# Polarization entanglement generation at 1.5 μm based on walk-off effect due to fiber birefringence

Qiang Zhou,[1] Wei Zhang,[1] Pengxiang Wang,[1] Yidong Huang,[1] and Jiangde Peng[1]

[1]*State Key Laboratory on Integrated Optoelectronics, Tsinghua National Laboratory for Information Science and Technology*
*Department of Electronic Engineering, Tsinghua University, Beijing, 100084, P. R. China*
*Email: betterchou@gmail.com*

In this Letter, a linear scheme to generate polarization entanglement at 1.5 μm based on commercial polarization maintained dispersion shifted fiber (PM-DSF) is proposed. The birefringent walk-off effect of the pulsed pump light in the PM-DSF provides an effective way to suppress the vector scattering processes of spontaneous four wave mixing. A 90 degree offset of fiber polarization axes is introduced at the midpoint of the fiber to realize the quantum superposition of the two correlated photon states generated by the two scalar processes on different fiber polarization axes, leading to polarization entanglement generation. Experiments of the indistinguishable property on single side and two-photon interference in two non-orthogonal polarization bases are demonstrated. A two photon interference fringe visibility of $89\pm3\%$ is achieved without subtracting the background counts, demonstrating its great potential in developing highly efficient and stable fiber based polarization-entangled quantum light source at optical communication band. © 2012 Optical Society of America
*OCIS Codes: 270.0270, 190.4380*

Polarization entangled photon pair source at 1.5 μm has important applications in quantum communication [1-3] and quantum information processing [4]. In recent years, spontaneous four wave mixing (SFWM) process in optical fiber has focused much attention as an important way to realize all-fiber quantum light source. When pulsed pump light passes through a piece of optical fiber, two kinds of SFWM processes would take place simultaneously [5]. One is the scalar scattering process, in which the two annihilated pump photons and the generated photon pairs are all polarized along the same fiber polarization axis. The other is the vector scattering process, in which the two annihilated pump photons are polarized along different fiber polarization axes, so do the generated photon pairs. For fibers without birefringence, such as dispersion shifted fibers (DSFs) employed in most previous experiments, the two processes are indistinguishable. The generated photon pairs always have the same polarization state with the pump light. To realize polarization entanglement, sophisticated schemes such as the time-multiplexing [6,7] and the polarization diversity loop [8,9], have been demonstrated to generate two orthogonally polarized correlated photon states and make them in quantum superposition in space and time.

Recently, utilizing the difference between the phase-matching conditions of the scalar and the vector scattering processes in a piece of micro-structured fiber (MSF) with group birefringence, we have realized polarization entangled photon pair generation employing a simple linear scheme [10]. These results demonstrate that the fiber birefringence could provide a much easier way to generate polarization entangled photon pairs. However, due to the lack of uniformity in performance, high splicing loss and high cost of the MSF, the potential on real application of the MSF based scheme is limited.

In this paper, we propose and demonstrate a new scheme for polarization entangled photon pair generation based on polarization maintained DSF (PM-DSF). When linearly polarized pulsed pump light injects into a piece of PM-DSF with a polarization direction of 45 degree respecting to fiber polarization axes, the two pump components polarized along the two polarization axes will walk off rapidly due to the high birefringence of the PM-DSF as shown in Fig. 1 (a), resulting an effective suppression of the vector scattering processes. Hence, correlated photon pairs are mainly generated by the two scalar scattering processes along the two fiber polarization axes. The corresponding two correlated photon states for the two scalar scattering processes are $|H_s\rangle|H_i\rangle$ and $|V_s\rangle|V_i\rangle$ respectively, where $H$ and $V$ denote two polarization directions along fiber axes, $s$ and $i$ denote signal and idler photons. However, $|H_s\rangle|H_i\rangle$ and $|V_s\rangle|V_i\rangle$ are not in quantum superposition due to the walk-off effect between fiber polarization axes.

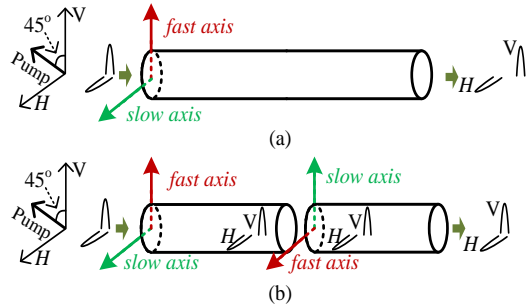

Fig. 1. (Color Online) The scheme of polarization entangled photon pair generation in the PM-DSF

To realize polarization entangled photon pair generation, the PM-DSF is split into two pieces with equal length and spliced together with 90 degree offset of fiber polarization axes, shown in Fig. 1 (b). Under this condition, two pump components along polarization axes will walk off rapidly in the first fiber section, and then walk together at the end of the second fiber section. The vector scattering process still can be suppressed effectively, thanks to the two pump components walk off entirely in the most part of the fiber. The $|H_s\rangle|H_i\rangle$ and $|V_s\rangle|V_i\rangle$ generated by the two scalar scattering processes overlap in space and time, since the walk-off effects in the two fiber sections cancel each other at the output end of the fiber. On the other hand, although the optical path in fiber is sensitive to environment, but the optical path variations of the two pump polarization components along $H$ and $V$ can also cancel each other in proposed scheme. Hence, the phase difference between $|H_s\rangle|H_i\rangle$ and $|V_s\rangle|V_i\rangle$, denoted by $\varphi$, is stable at the output end of the PM-DSF, ensuring the quantum superposition of the two correlated photon states and realizing the generation of the polarization entangled state of $1/\sqrt{2}\left(|H_s\rangle|H_s\rangle + e^{i\varphi}|V_s\rangle|V_i\rangle\right)$.

To demonstrate the vector scattering processes suppression in the scheme shown in Fig. 1 (b), the photon flux spectral densities (PFSDs) of correlated photon pairs generated by the scalar and the vector scattering processes are compared theoretically utilizing the expressions deduced in Ref. [5], as shown in Eq. (1) and (2)

$$f_{HH} = (\gamma P_H L_s)^2 \operatorname{sinc}^2\left[\left(\beta_2 \Omega^2 + 2\gamma P_H\right)\frac{L_s}{2}\right]$$
$$f_{VV} = (\gamma P_V L_s)^2 \operatorname{sinc}^2\left[\left(\beta_2 \Omega^2 + 2\gamma P_V\right)\frac{L_s}{2}\right]$$
(1)

$$f_{HV} = \frac{4}{9}\left(\gamma \sqrt{P_H P_V} L_v\right)^2 \times$$
$$\operatorname{sinc}^2\left[\left(\Delta\beta_1 \Omega + \beta_2 \Omega^2 + \gamma(P_H + P_V)\right)\frac{L_v}{2}\right]$$
$$f_{VH} = \frac{4}{9}\left(\gamma \sqrt{P_H P_V} L_v\right)^2 \times$$
$$\operatorname{sinc}^2\left[\left(\Delta\beta_1 \Omega - \beta_2 \Omega^2 - \gamma(P_H + P_V)\right)\frac{L_v}{2}\right]$$
(2).

Where, $f_{HH}$ and $f_{VV}$ are the PFSDs of two scalar scattering processes along $H$ and $V$, respectively. $f_{HV}$ and $f_{VH}$ are the PFSDs of the two vector scattering processes in which the generated signal and idler photons are polarized along $H$ and $V$ or $V$ and $H$, respectively. $P_H$ and $P_V$ are the power of two pump components polarized along $H$ and $V$, respectively. $\Omega$ is the frequency detuning between pump wavelength and the wavelength of the signal or idler photons. $\gamma$ is the optical nonlinear coefficient of the PM-DSF. $\Delta\beta_1$ and $\beta_2$ are group birefringence and group velocity dispersion of the PM-DSF at the pump wavelength, respectively. $L_s$ and $L_v$ are the effective length in which the scalar and the vector scattering processes could take place.

Parameters used in theoretical analysis refer to the experimental setup, where linearly polarized pump pulses with a pulse width of about 20 ps are injected into the PM-DSF (DS15-PS-U40A, fabricated by Fujikura Ltd.) with 90 degree offset of fiber polarization axes at the midpoint. $L_s$ is 150 m, which is the total length of the fiber exactly. $\Delta\beta_1$ of the PM-DSF is 0.286 ps/m, hence the two pump polarization components would walk off entirely after a transmission of 7.5 m at the input end. Since the two pump polarization components would walk together at the output end of the second fiber section, $L_v$ is about 15 m, i.e. double of the walk-off length. $\beta_2$ and $\gamma$ are -3.824 ps²/km and 3 /W/km for PM-DSF used in experiment. The total pump peak power is defined as $P_p$ and the angle between the polarization direction of the pump light and $H$ is defined as $\theta$. The peak powers of the two pump polarization components, $P_H$ and $P_V$ could be obtained by $P_p \cos^2\theta$ and $P_p \sin^2\theta$, respectively.

The PFSDs of generated photon pairs by the scalar and the vector scattering processes under $P_p = 0.8$ W and $\theta = 45$ degree are shown in Fig. 2. The solid line in the main figure of Fig. 2 is the PFSD of the total scalar scattering processes. The dashed line in the inset of Fig. 2 is the PFSD of the total vector scattering processes, respectively. It can be seen that both scalar and vector scattering processes can generate photon pairs near the pump wavelength. However, the PFSD of the vector scattering processes is much lower than that of the scalar scattering processes due to $L_v \ll L_s$. The filters for the signal and idler photons in the experiment setup have a frequency detuning of 0.2 THz, which is indicated by the dash-dotted line in Fig. 2. It shows that under the frequency detuning of 0.2 THz the scalar scattering processes is about two orders of magnitude greater than the vector scattering processes. It demonstrates that the walk-off effect in the PM-DSF can provide sufficient suppression of the vector scattering processes, which is the essential of the scheme to realize polarization entanglement generation.

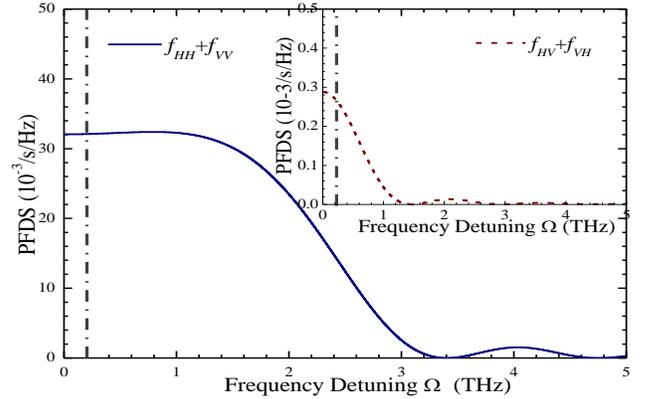

Fig. 2. (Color Online) PFSD of generated photon pairs by the scalar and the vector scattering processes; the main figure: $f_{HH} + f_{VV}$; the inset: $f_{HV} + f_{VH}$

The experimental setup for the proposed scheme is shown in Fig. 3, which is based on commercial components for optical communication. The pulsed pump is generated by a passive mode-locked fiber laser with a central wavelength of 1552.75 nm, a pulse width of about 20 ps and a repetition rate of 1 MHz. A side-band rejection of >115 dB is achieved at the wavelengths of signal photons and idler photons. The polarization state of pump is controlled by a polarizer (P), a rotatable half wavelength plate (HWP1) and a polarization controller (PC1). A variable optical attenuator (VOA) and a 50/50 fiber coupler with a power meter (PM) are used to adjust and monitor the pump power. The PM-DSF is 150 m in length, submerged in liquid nitrogen (77 K) to suppress the impact of spontaneous Raman scattering (SpRS) process. The PM-DSF is divided into two pieces with equal length and then spliced together with 90 degree offset of fiber polarization axes. The output of PM-DSF is directed into a filtering and splitting system based on a 100 GHz/40-channels arrayed waveguide grating (AWG, Scion Photonics Inc.), two fiber Bragg gratings (FBGs) and two tunable optical band-pass filters (TOBFs). The total pump isolation is >110 dB at either signal (1555.15 nm) or idler wavelength (1550.35 nm). Two single photon detectors (SPDs, Id Quantique, id201) are operated under Geiger mode with a detection window of 2.5 ns, which are triggered with the residual pump detected by a photo-detector (PD). The detection efficiency of SPD1 and SPD2 are 21.83% and 22.56%, respectively.

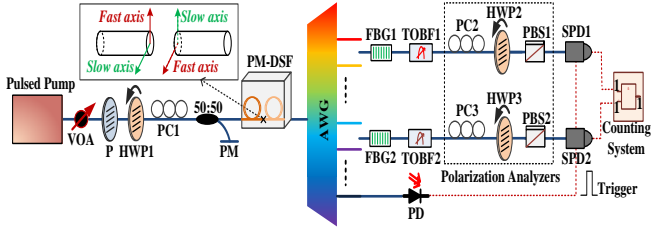

Fig. 3. (Color Online) Experimental setup

In the experiment $\theta$ is adjusted to 45 degree for the maximally polarization entangled photon pair generation. To measure the polarization entanglement of generated photon pairs, two polarization analyzers are inserted before SPDs, shown in the dashed square in Fig. 3. Each of them consists of a PC, a rotatable HWP and a polarizing beam splitter (PBS). The polarization direction of the two polarization analyzers are adjusted through the method mentioned in [10], the detecting polarization direction of signal and idler side are denoted by $\theta_s$ and $\theta_i$, which are adjusted by rotating the HWP2 and the HWP3, respectively.

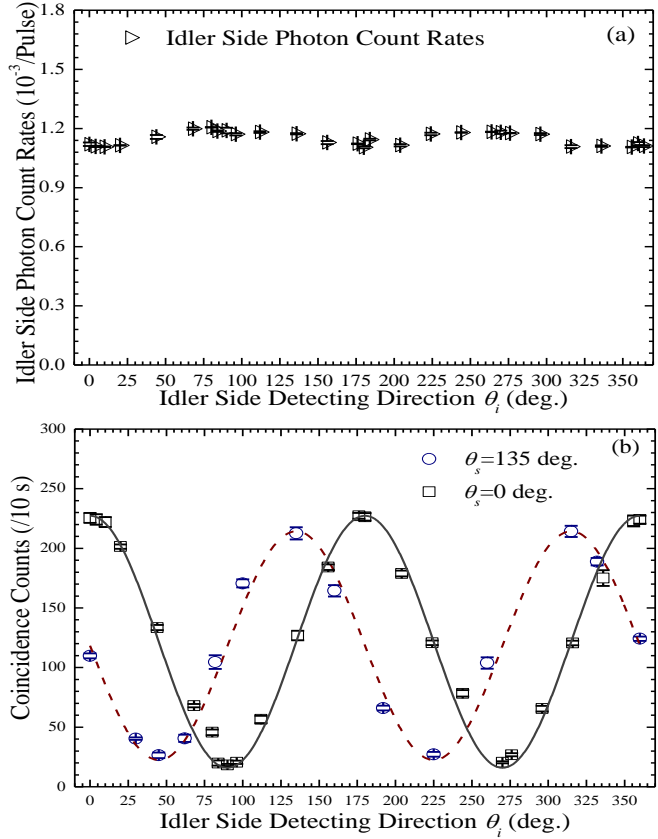

Fig. 4. (Color Online) (a) Idler side photon count rates under different $\theta_i$; (b) Coincidence counts under different $\theta_i$.

Figure 4 shows the measured entanglement properties of generated photon pairs. The idler side photon count rates under different $\theta_i$ are shown in Fig. 4 (a), which are almost unchanged with $\theta_i$, except a small ripple caused by the polarization dependent loss of the HWP3, demonstrating polarization indistinguishable property of the generated polarization entangled photon pairs. The single count rate is about $10^{-3}$ per pulse, ensuring that the possibility of multi-photon pair event is sufficient low in the experiment. Fig. 3 (b) is the results of coincidence counts per 10 seconds (without subtracting the background count) under different $\theta_i$. Squares and circles are experimental data while $\theta_s$ is set to 0 and 135 degree, respectively. Solid and dashed lines are fitting curves, showing that the visibilities of two-photon interference (TPI) fringes are 92±3% and 89±3% for $\theta_s$ = 0 and 135 degree respectively. Hence, the proposed scheme of polarization entangled photon pair generation based on commercial PM-DSF is demonstrated by the experiment results of the indistinguishable property on single side and TPI in two non-orthogonal polarization bases.

In summary, a linear scheme to generate polarization entangled photon pairs at 1.5 μm based on commercial PM-DSF is proposed and demonstrated experimentally. The birefringent walk-off effect in the PM-DSF provides effective way to suppress the vector scattering processes in the fiber. A 90 degree offset of fiber polarization axes is introduced at the midpoint of the fiber to realize the quantum superposition of the two correlated photon states generated by the two scalar processes along fiber polarization axes. A TPI fringe visibility of 89±3% is achieved without subtracting the background counts, showing the great potential of the proposed scheme in developing efficient and stable fiber based polarization entangled quantum light source at optical communication band.

This work is supported in part by 973 Programs of China under Contract No. 2011CBA00303 and No. 2010CB327606, and Basic Research Foundation of Tsinghua National Laboratory for Information Science and Technology (TNList).